\documentclass[12pt]{iopart}

\usepackage{iopams}  

\usepackage{graphicx}
\usepackage{color}

\begin{document}

\def\tr{\text{tr}}
\def\Tr{\mbox{Tr}}

\newtheorem{theo}{Theorem}

\title{Ergotropy as distance from equilibrium}
\author{Michele Campisi}
\address{Istituto Nanoscienze-CNR, NEST-Scuola Normale Superiore, 56126 Pisa, Italy}
\date{\today }

\begin{abstract}
We show that the ergotropy (a.k.a available energy) of a thermally isolated system is the $1$-Wasserstein distance between two energy probability density functions. Specifically, between the energy profile of the system state, and the energy profile of its associated passive state. We prove this for both classical and quantum systems, and illustrate it with examples.  Our result formalises the intuition that ergotropy is a quantifier of distance from equilibrium 
\end{abstract}

\maketitle

\section{Introduction}
The ergotropy, a.k.a. available energy, of a thermally isolated system is the maximal amount of energy it can release as a consequence of an external time dependent driving \cite{Allahverdyan04EPL67,Campisi26EPL154,Hatsopoulos76FP6b,Gardner63PF6,Helander17JPP83}. 
After the extraction of ergotropy a system remains in the so-called passive state $\breve \rho$ associated to the initial state $\rho$ \cite{Lenard78JSP19,Pusz78CMP58,Gorecki80LMP4,Daniels81JMP22}. The passive state is energetically inert, namely it is such that it cannot release any energy if externally perturbed: in fact it can only absorb energy. The most prominent example is the thermal equilibrium state of canonical form $e^{-\beta H}/Z$. However the family of passive states is much larger and includes all states that are stationary and are characterised, just like the canonical ensemble, by the fact that the probability diminishes as the energy increases. In this regard passive sates can be understood as generalised thermodynamic equilibria.

Intuitively, the higher the ergotropy the more one expects the system to be away from its associated equilibrium state. This intuition was strongly corroborated by the recent work of Ref. \cite{Mackenbach22PRL128} highlighting that the available energy of the electrons trapped in a flux tube of a collisionless plasma is highly correlated to the amount of turbolent energy transport. 

Can this intuition be formalised? Namely, can ergotropy be expressed, mathematically, as some ``distance''  from equilibrium?  Here we answer in the positive.

To do so we single out the appropriate space and the proper notion of distance over said space, which coincides with the notion of ergotropy.
The space is that of  energy probability density functions, and the distance is the $1$-Wasserstein distance over said space.
Specifically, we prove that given a state $\rho$, its ergotropy is the $1$-Wasserstein distance between its energy profile, and the energy profile of its passive companion $\breve \rho$, namely the according equilibrium. 

We prove this result for both classical and quantum systems and illustrate it with the 1D harmonic oscillator, both classical and quantum.

\section{Main result: classical case}

Given a classical system with Hamiltonian $H(\mathbf z)$, being in a statistical state described by the probability density $\rho(\mathbf z)$ (here $\mathbf z=(\mathbf q , \mathbf p)$ is a point in the $f$-dimensional phase space), we ask the maximal amount of energy that can be extracted from the system by means of a cyclic variation of an external parameter.

That is, we seek the quantity
\begin{eqnarray}
\mathcal E[\rho] =  \int d\mathbf z \rho(\mathbf z) H(\mathbf z) -\min_\phi \int d\mathbf z \rho(\mathcal \phi^{-1}(\mathbf z)) H(\mathbf z)
\end{eqnarray}
where the minimisation is carried over all symplectic diffeomorphisms $\phi$ in phase space.
$\mathcal E$ is known as the Ergotropy of $\rho$, relative to $H$. This is the classical analogue of the well-known ergotropy of a quantum system \cite{Allahverdyan04EPL67}. In atmospheric, plasma- and astro-physics it is known as ``available energy'' \cite{Gardner63PF6,Helander17JPP83,Lemou12IM187}.

The state $\breve \rho$ that reaches the minimal energy, known as the ground state, or the passive companion of $\rho$, is mathematically speaking a measure preserving rearrangement of $\rho$, specifically the rearrangement according to which the probability density  $\breve \rho$ does not increase as one moves in phase space from one  point to a point of higher energy $H$ \cite{Campisi26EPL154,Campisi26arXiv:2603.28388}. Such ``ergotropic rearrangement'' has been shown recently \cite{Campisi26EPL154,Campisi26arXiv:2603.28388}, to be given by the compact expression
\begin{equation}
\breve \rho = \Sigma^{-1}\circ \Omega \circ H
\end{equation}
where 
\begin{eqnarray}
\Sigma(r)&=  \int d\mathbf z \, \theta[\rho(\mathbf{z})-r]
\label{eq:Sigma}\\
\Omega(E) &= \int d\mathbf z \, \theta[E-H(\mathbf{z})]
\label{eq:Omega_0}
\end{eqnarray}
and 
\begin{eqnarray}
\Sigma^{-1}(y)= \int_0^\infty dx\, \theta [\Sigma(x)-y] 
\end{eqnarray}
 is the (generalized) inverse of $\Sigma$ \footnote{This definition extends the standard notion of the inverse of a strictly decreasing continuous function, to the case of a non-increasing function possibly featuring discontinuous jumps and plateaus.}.
The symbol $\theta(\cdot)$ denotes the Heaviside step function.

Note that, given $H$, $\breve \rho$ is uniquely determined by $\rho$, and that $\breve \rho$ is a stationary state, because its Poisson brakets with $H$ are trivially null. 
 Note that by definition $\Omega$ is non-decreasing (in fact for any physical Hamiltonian $\Omega$ is strictly increasing).
Similarly, $\Sigma$ is non-increasing, so $\breve \rho$ is non-increasing as $H$ increases. 
 
The ergotropy of the state $\rho$ is thus given by the difference between the energy expectation of $\rho$ and that of $\breve \rho$
\begin{eqnarray}
\mathcal E[\rho] = \int d\mathbf z H(\mathbf z) [\rho(\mathbf z)- \breve \rho(\mathbf z)] 
\end{eqnarray}

In order to state our main result we need the following two ingredients.

First, let us introduce the energy profile $\mathcal P(E)$ of $\rho$, namely the probability density that the system lies on the hyper-surface $H(\mathbf z)=E$, given that it is distributed in phase space according to $\rho$, namely,
\begin{eqnarray}
\mathcal P(E)=\int d\mathbf z {\delta [E-H(\mathbf z)]} \rho(\mathbf z)
\label{eq:mathcalP}
\end{eqnarray}
Similarly, let  $\breve \mathcal P(E)$ be the energy profile of $\breve \rho$
\begin{eqnarray}
\breve \mathcal P(E)=\int d\mathbf z {\delta [E-H(\mathbf z)]} \breve \rho(\mathbf z)
\end{eqnarray}

Second, let us introduce the $1$-Wasserstein distance $W_1[Q_1,Q_2]$ between two normalized probability distributions, $Q_i:\mathbb R^N\to R_+, i=1,2$
\begin{eqnarray}
W_1[Q_1,Q_2] = \inf_{\gamma \in \Gamma} \int dx dy\,  \gamma(x,y) |x-y|
\label{eq:$1$-Wasserstein0}
\end{eqnarray}
where $\Gamma$ is the set of all joint probability distributions $\gamma(x,y)$ (a.k.a. ``transport plans'' in this context) whose marginals are $Q_1(x)$ and $Q_2(y)$, namely $\int dx \gamma(x,y)=Q_2(y), \int dy \gamma(x,y)=Q_1(x)$. $D$ is a proper metric on the space of distributions. 

The minimisation problem (\ref{eq:$1$-Wasserstein0}) has a simple analytical solution in the case of probability distributions on the line ($N=1$) \cite{Vallender74TPA18}:
\begin{eqnarray}
W_1[Q_1,Q_2] = \int dx |c_1(x)-c_2(x)| 
\label{eq:$1$-Wasserstein}
\end{eqnarray}
where 
\begin{eqnarray}
c_i(x)=\int^x_{-\infty} ds \, Q_i(s)
\end{eqnarray}
denote the cumulative distribution functions.

Our main result is the following

\begin{theo}
The ergotropy of $\rho$ is the $1$-Wasserstein distance between the energy profile of $\rho$, and the energy profile of $\breve \rho$
\begin{eqnarray}
\mathcal E[\rho] = W_1[\mathcal P,\breve \mathcal P]
\end{eqnarray}
\end{theo}

Using the energy profiles $\mathcal P,\breve{\mathcal P}$, the ergotropy can be readily written as

\begin{eqnarray}
\mathcal E =  \int_0^\infty dE\, [\mathcal P(E)- \breve \mathcal P(E)] E\,. \label{eq:c-ergotropy2}
\end{eqnarray}
where we have conventionally set the minimum of $H$ to $0$.
Introducing the cumulative probability functions 
\begin{eqnarray}
\mathcal C(x) &=  \int_0^x dE\, \mathcal P(E) \label{eq:cumulative-cl}\\
\breve \mathcal C(x) &=  \int_0^x dE\, \breve \mathcal P(E) 
\end{eqnarray}
and integrating by parts, we get:
\begin{eqnarray}
\mathcal E =  \left[(\mathcal C(x)-\breve \mathcal C(x))x \right]_0^\infty - \int_0^\infty dE\, [\mathcal C(E)-\breve \mathcal C(E)] \label{eq:ergotropy3}
\end{eqnarray}
Note that, in the limit $x\to \infty$, it is $\mathcal C(x)\to 1, \breve \mathcal C(x)\to 1$, therefore $\mathcal C(x)-\breve \mathcal C(x) \to 0$, but  the product $(\mathcal C(x)-\breve \mathcal C(x) )x$ is indeterminate. However, because $\mathcal P,\breve \mathcal P$ have finite first  moment (because we assumed from the start that $\rho,\breve \rho$ have a well defined energy expectation), it means they go to zero faster than $1/x^2$ for large $x$. Thus, using de l'H\^opital rule, we get
\begin{eqnarray}
&\lim_{x\to \infty} [\mathcal C(x)-\breve \mathcal C(x)]x
=\lim_{x\to \infty} \frac{\breve \mathcal C(x)-\mathcal  C(x)}{1/x} \\
&= \lim_{x\to \infty} \frac{\mathcal P(x)-\breve \mathcal P(x)}{1/x^2} 
=\lim_{x\to \infty} \frac{\mathcal P(x)}{1/x^2}-  \lim_{x\to \infty} \frac{\breve \mathcal P(x)}{1/x^2}
=0
\end{eqnarray}
On the other hand, trivially $(\mathcal C(x)-\breve \mathcal C(x))x \to 0$, for $x\to 0$, therefore
\begin{eqnarray}
\mathcal E =  \int_0^\infty dE\, [\breve \mathcal C(E)-\mathcal C(E)] \label{eq:ergotropy4}
\end{eqnarray}

Next, let's consider the probability density $P(\Phi)$ ($\breve P(\Phi)$) that the system is on the surface of constant energy that encloses the volume $\Phi$, provided it is distributed in phase space according to $\rho (\breve \rho)$, namely \cite{Campisi26EPL154}
\begin{eqnarray}
P(\Phi) &=  \int d\mathbf z  \delta[\Phi-\Omega(H(\mathbf z))] \rho(\mathbf z)\\
\breve P(\Phi)&= \int d\mathbf z \delta[\Phi-\Omega(H(\mathbf z))]  \breve \rho (\mathbf z) \label{eq:breveP}
\end{eqnarray}

For simplicity we shall call $P, \breve P$, the phase-volume-profiles of $\rho,\breve \rho$, respectively.
We  recall that, as shown in Ref. \cite{Campisi26EPL154}, 
\begin{eqnarray}
P(\Phi) &=  \int d \Theta G[\Theta|\Phi] \breve P(\Theta)\,, \label{eq:P0-P1}
\end{eqnarray}
where $G$ is the overlap function
\begin{eqnarray}
G[\Theta|\Phi] &= \int d\mathbf z \delta[\Phi-\Omega(H(\mathbf z))]  \delta[\Theta-\Sigma(\rho(\mathbf z))] 
\label{eq:G}
\end{eqnarray}
We remark that  $G$ is a ``doubly stochastic'' conditional probability density \cite{Campisi25Book}, namely
\begin{eqnarray}
\int d\Theta G[\Theta|\Phi] = \int d\Phi G[\Theta|\Phi] =1, \quad G\geq 0
\label{eq:doubly-stoc}
\end{eqnarray}
This implies that (see Ref. \cite{Campisi25Book}, pag 99)
\begin{eqnarray}
\int_0^x d\Phi P(\Phi) \leq \int_0^x d\Phi \breve P(\Phi)\, \quad \forall x\geq 0 
\label{eq:P-majorisation}
\end{eqnarray}

 It is not difficult to see that energy and phase-volume profiles are linked by the change of variable $E\to \Phi =\Omega(E)$:
\begin{eqnarray}
\mathcal P(E) &=  P(\Omega(E)) \omega(E) 
\end{eqnarray}
where
\begin{eqnarray}
\omega(E)=\frac{d}{dE}\Omega(E)
\end{eqnarray}
is the density of states  \cite{Campisi25Book,Khinchin49Book}.
In fact, we have:
\begin{eqnarray}
 P(\Omega(E)) &=  \int d\mathbf z  \delta[\Omega(E)-\Omega(H(\mathbf z))] \rho(\mathbf z)\\
&=  \int d\mathbf z  \frac{\delta[E-H(\mathbf z)]}{\omega(E)} \rho(\mathbf z)=  \frac{\mathcal P(E)}{\omega(E)}
\end{eqnarray}
where we used $\delta(f(x)-a)=\delta(x-f^{-1}(a))/|f'(f^{-1}(a))|$ for a strictly increasing function $f$. Similarly, 
\begin{eqnarray}
\breve \mathcal P(E)  &=  \breve P(\Omega(E)) \omega(E)
\end{eqnarray}

Therefore, choosing $x=\Omega(y)$ as the upper integration limit in Eq. (\ref{eq:P-majorisation}), and making the usual change of variable $\Phi \to E$, we get:
\begin{eqnarray}
\int_0^y dE \mathcal P(E) \leq \int_0^y dE \breve \mathcal P(E)\, \quad \forall y\geq 0 
\end{eqnarray}
that is 
\begin{eqnarray} 
\mathcal C(y)\leq  \breve \mathcal C(y)\, .
\label{eq:C<breveC}
\end{eqnarray}
This allows to replace  $[\breve \mathcal C(E)-\mathcal C(E)]$ with $| \breve \mathcal C(E)-\mathcal C(E)|$ in Eq. (\ref{eq:ergotropy3}), and obtain
\begin{eqnarray}
\mathcal E =  \int_0^\infty dE\, | \breve \mathcal C(E)-\mathcal C(E) | \label{eq:ergotropy4}
\end{eqnarray}
Noting that the r.h.s. of the above equation is the $1$-Wasserstein distance between $\mathcal P$ and $\breve \mathcal P$, Eq. (\ref{eq:$1$-Wasserstein}), concludes the proof.

\section{Main result: quantum case}
In the quantum case, the ergotropy is given by 
\begin{eqnarray}
\mathcal E[\hat\rho] = \Tr  \hat \rho \hat H - \min_{\hat U} \Tr \hat U \hat \rho \hat U^\dagger
\end{eqnarray}
where the minimisation is over all unitary operators $U$. The passive companion of $\hat \rho$, $\breve {\hat \rho}$, is known to be given by the expression \cite{Allahverdyan04EPL67}: 
\begin{eqnarray}
\breve {\hat \rho}= \sum_i r_i |{E_i}\rangle \langle{E_i}|
\end{eqnarray}
where $|E_i\rangle$ denote the eigenvectors of $\hat H$, and we adopted the convention that the eigenvalues $r_i$ of $\rho$ are ordered in non-increasing fashion while the eigenvalues, $E_n$, of $H$ are ordered in non-decreasing fashion, 
\begin{eqnarray}
E_0& \leq E_1 \leq \dots      \label{eq:energies_ordered}\\
r_0 & \geq r_1 \geq \dots     \label{eq:populations_ordered}
\end{eqnarray}

Accordingly, the ergotropy can be swiftly expressed as:
\begin{eqnarray}
\mathcal E[\hat\rho] = \sum_i (p_i -r_i) E_i
\end{eqnarray}
where
\begin{eqnarray}
p_i = \sum_{j} |\langle{E_i}|r_j\rangle|^2 r_j
\label{eq:bistoch-quantum}
\end{eqnarray}
with $|r_i\rangle$ denoting the eigenvectors of $\hat \rho$.

The energy profiles of $\hat \rho$, $\breve{\hat \rho}$ are given, respectively, by
\begin{eqnarray}
\mathcal P (E) = \sum_i p_i \delta(E-E_i)\label{eq:P-quantum}\\
\breve \mathcal P (E) = \sum_i r_i \delta(E-E_i) \label{eq:brevePquantum}
\end{eqnarray}
They represent the probability density that a projective measurement of the observable $H$ has outcome $E$ provided the system is, respectively, in the state $\hat \rho$, $\breve{\hat \rho}$. Note the analogy with the classical case, where an integral of Dirac deltas gets replaced by a discrete sum of Dirac deltas.

The following holds:

\begin{theo}
The ergotropy of $\hat \rho$ is the $1$-Wasserstein distance between the energy profile of $\hat \rho$, and the energy profile of $\breve{\hat \rho}$
\begin{eqnarray}
\mathcal E[\hat \rho] = W_1[\mathcal P,\breve \mathcal P]
\end{eqnarray}
\end{theo}

Using the energy profiles, Eqs. (\ref{eq:P-quantum},\ref{eq:brevePquantum}), the ergotropy reads exactly like in the classical case, i.e., as:
\begin{eqnarray}
\mathcal E =  \int_0^\infty dE\, [\mathcal P(E)- \breve \mathcal P(E)] E\,. \label{eq:q-ergotropy2}
\end{eqnarray}
Performing the integration by parts as in the classical case, one gets to 
\begin{eqnarray}
\mathcal E =  \int_0^\infty dE\, [\breve \mathcal C(E)- \mathcal C(E)]
\end{eqnarray}
where 
\begin{eqnarray}
\mathcal C (E) = \sum_i p_i \theta(E-E_i)\label{eq:C-quantum}\\
\breve \mathcal C (E) = \sum_i r_i \theta(E-E_i) \label{eq:breveCquantum}
\end{eqnarray}
are the cumulative probability functions associated to $\mathcal P$ and $\breve \mathcal P$.

Since the probabilities $p_i$ are linked to the probabilities $r_j$ by a bi-stochastic matrix, whose elements are the quantum overlaps $|\langle{E_i}|r_j\rangle|^2$, see Eq. (\ref{eq:bistoch-quantum}), one can prove (see appendix) that $\breve \mathcal C (E) \geq \mathcal C(E)$. This allows to replace $\breve \mathcal C(E)- \mathcal C(E)$ with $|\breve \mathcal C(E)- \mathcal C(E)|$, in Eq. (\ref{eq:q-ergotropy2}), which concludes the proof.

\section{Illustrative examples}
\subsection{Classical harmonic oscillator}
Consider a 1D classical harmonic oscillator
\begin{eqnarray}
H(q,p)= \frac{p^2}{2}+\frac{q^2}{2}
\end{eqnarray}
being in the Gaussian state centered at $(q,p)=(a,0)$:
\begin{eqnarray}
\rho(q,p)= \frac{1}{2\pi \sigma }e^{-\frac{(q-a)^2+p^2}{2\sigma}}
\label{eq:gaussian-state}
\end{eqnarray}
Its ergotropic rearrangement is 
\begin{eqnarray}
\breve \rho(q,p)= \frac{1}{2\pi \sigma }e^{-\frac{q^2+p^2}{2\sigma}}
\end{eqnarray}
To see that note that $\rho$ can be mapped onto $\breve \rho$ via a translation, which is a measure preserving transformation. Specifically, 
$\breve \rho(q,p)=\rho(\mathcal M^{-1}(q,p))$, where $\mathcal M(q,p)=(q-a,p)$. Furthermore, $\breve \rho(q,p)$ is a decreasing function of the energy, thus it is the passive companion of $\rho$.

The system energy is easily calculated by means of Gaussian integrals:
\begin{eqnarray}
e = \frac{1}{2\pi \sigma } \int dq dp \left(\frac{p^2}{2}+\frac{q^2}{2}\right) e^{-\frac{(q-a)^2+p^2}{2\sigma}} = \frac{a^2}{2}+\sigma
\end{eqnarray}
Similarly the passive state energy expectation is given by
\begin{eqnarray}
\breve e = \frac{1}{2\pi \sigma } \int dq dp \left(\frac{p^2}{2}+\frac{q^2}{2}\right) e^{-\frac{q^2+p^2}{2\sigma}} = \sigma
\end{eqnarray}
Their difference $e-\breve e$ gives the ergotropy:
\begin{eqnarray}
\mathcal E = \frac{a^2}{2}
\end{eqnarray}
Note the interesting fact that it does not depend on the width $\sigma$ of the Gaussian state. Thus the ergotropy is, so to speak, simply the energy $H(q,p)$ of the center of mass $(q,p)=(a,0)$ of the Gaussian distribution.

Let us calculate the energy profile $\mathcal P$ of the state $\rho$. Using the definition in Eq. (\ref{eq:mathcalP}), we have:
\begin{eqnarray}
\mathcal P(E) = \int dq dp {\delta(E-p^2/2-q^2/2)}  \frac{1}{2\pi \sigma }e^{-\frac{(q-a)^2+p^2}{2\sigma}}
\end{eqnarray}
To perform the integral we use the celebrated formula \cite{Khinchin49Book,Campisi25Book}:
\begin{eqnarray}
\int d\mathbf z \delta(E-H(\mathbf z)) f(\mathbf z) = \int_{E=H} \frac{d S}{|\nabla H|} f(\mathbf z)
\end{eqnarray}
where $\int_{E=H} d S$ denotes integration over the surface $H=E$. In the case of the harmonic oscillator the surface $H=E$ is simply the circle of radius $R=\sqrt{2E}$. Thus $dS = R d\phi=\sqrt{2E} d\phi $, where $\phi$ is the angular coordinate. The gradient of $H$ is $\nabla H= (q,p)$, thus $|\nabla H|=\sqrt{q^2+p^2}=\sqrt{2E}$, on the $H=E$ surface. Using  $q=\sqrt{2E} \cos \phi, p=\sqrt{2E}\sin \phi$ we then obtain:
\begin{eqnarray}
\mathcal P(E)  &=  \frac{e^{-E/\sigma }}{2 \pi \sigma} e^{-\frac{a^2}{2\sigma}} \int_0^{2\pi} d\phi e^{\frac{a\sqrt{2E}}{\sigma}\cos \phi} \\
&=  \frac{e^{-E/\sigma }}{\sigma} e^{-\frac{a^2}{2\sigma}} I_0 ({a\sqrt{2E}}/{\sigma})
\end{eqnarray}
where $I_0(x)$ is the modified Bessel function of the first kind \cite{Gradshteyn07Book}. 

The ground state energy profile $\breve \mathcal P$ is obtained by setting $a=0$ in the above expression of $\mathcal P$, which gives:
\begin{eqnarray}
\breve \mathcal P(E)  =  \frac{e^{-E/\sigma }}{\sigma} 
\end{eqnarray}

\begin{figure}[t]
\includegraphics[width=.5\linewidth]{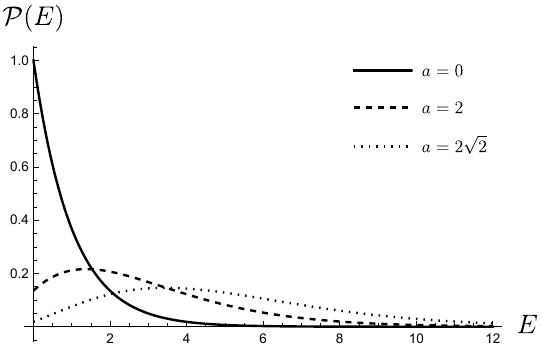}
\includegraphics[width=.5\linewidth]{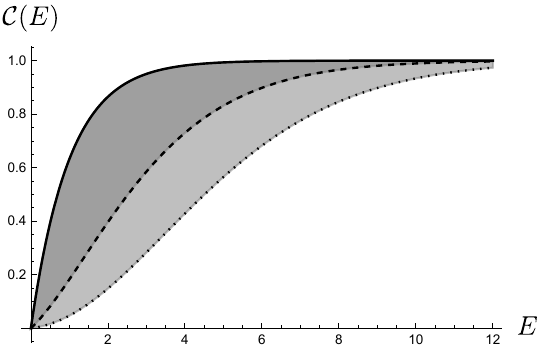}
\caption{Left Panel: Energy profiles, $\mathcal P(E)$, of Gaussian states, Eq. (\ref{eq:gaussian-state}), and of their passive companion ($a=0$), for a classical 1D harmonic oscillator. 
Right panel: according cumulative functions, $\mathcal C(E)$, Eq. (\ref{eq:cumulative-cl}). All graphs plotted for $\sigma=1$. }
\label{fig:PEc+CEc} 
\end{figure}

Figure \ref{fig:PEc+CEc}, left panel, shows plot of $\mathcal P(E)$ for $a=2$, $a=2\sqrt{2}$ and the according plot of $\breve{\mathcal P}(E)$, that is the $a=0$ energy profile. The chosen value of $\sigma$ is $\sigma=1$. In the right panel are displayed the according cumulative functions $\mathcal C(E)$. The total shaded area is the ergotropy of the $a=2\sqrt{2}$ state, amounting to $\mathcal E= 4$. The darker shaded area, similarly is the ergotropy of the $a=2$ state, amounting to $\mathcal E= 2$. Accordingly, the light shaded area is the energy difference $\Delta E$ between the states with $a=2\sqrt{2}$ and $a=2$, which amounts to $\Delta E=2$.

\subsection{Quantum harmonic oscillator}
Consider a 1D quantum harmonic oscillator:
\begin{eqnarray}
\hat H= \frac{\hat p^2}{2}+\frac{\hat q^2}{2}
\end{eqnarray}
being in the coherent state
\begin{eqnarray}
| \alpha \rangle = e^{-|\alpha|^2/2} \sum_{n=0}^{\infty} \frac{\alpha^n}{\sqrt{n!}} | E_n \rangle
\end{eqnarray}
where $| E_n \rangle$ denote the eigenstates of $\hat H$. The system's density operator is:
\begin{eqnarray}
\hat \rho = | \alpha \rangle \langle \alpha |
\end{eqnarray}
Since $\hat \rho$ is a pure state, its passive companion is simply the quantum mechanical ground state
\begin{eqnarray}
\breve{\hat \rho} = | E_0\rangle \langle E_0 |
\end{eqnarray}
The probability $p_n$ of finding the eigenenergy $E_n = \left( n+\frac{1}{2}\right) $ is  given by the Poisson distribution:
\begin{eqnarray}
p_n= e^{-|\alpha|^2} \frac{|\alpha|^{2n}}{n!}
\label{eq:pn-coherent}
\end{eqnarray}
The according energy expectation is 
\begin{eqnarray}
e = \sum_n p_n E_n = |\alpha|^2 + \frac{1}{2}
\end{eqnarray}
The probability $r_n$ of finding the eigenenergy $E_n $ in the ground state $\breve{\hat \rho}$ is trivially $1$ if $n=0$, and null otherwise:
\begin{eqnarray}
r_n= \delta_{0,n}
\end{eqnarray}
where $\delta_{a,b}$ is the Kronecker symbol. The according energy expectation is simply the ground state energy
\begin{eqnarray}
\breve e = \frac{1}{2}
\end{eqnarray}
Thus the ergotropy amounts to
\begin{eqnarray}
\mathcal E = |\alpha|^2 
\end{eqnarray}
\begin{figure}[t]
\includegraphics[width=.5\linewidth]{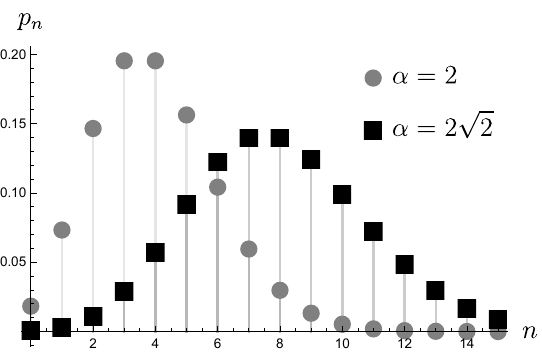}
\includegraphics[width=.5\linewidth]{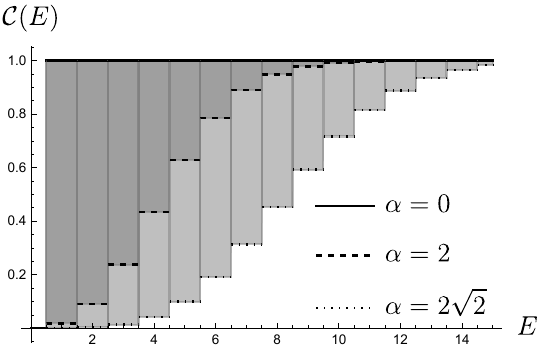}
\caption{Left Panel: Energy eigenstates populations, $p_n$, of coherent states, Eq. (\ref{eq:pn-coherent}), of a 1D quantum harmonic oscillator.
Right panel: according cumulative functions, $\mathcal C(E)$, Eq. (\ref{eq:C-quantum}), and the cumulative function $\breve{\mathcal C}(E)$ of the ground state (i.e. the $\alpha=0$ curve). }
\label{fig:PEq+CEq} 
\end{figure}
As shown in Fig. \ref{fig:PEq+CEq} the cumulative functions (\ref{eq:C-quantum},\ref{eq:breveCquantum}) are, in this specific case, histograms made of  bars of width $1$ (except for the $n=0$ bar, which is of width $1/2$).
Accordingly, the $1$-Wasserstein distance is easily written as
\begin{eqnarray}
W_1[\mathcal P, \breve{\mathcal P}] = \sum_{n=0}^{\infty}
\sum_{k=0}^{n} (r_k- p_k) = 
\sum_{n=0}^{\infty} \left[
1- e^{-|\alpha|^2} \sum_{k=0}^n  \frac{|\alpha|^{2k}}{k!}\right]
\label{eq:D-qHO}
\end{eqnarray}
As shown in the  appendix, the above sum amounts to $|\alpha|^2$, i.e., the ergotropy, in agreement with our main result.

\section{Remarks}
Given that the ergotropy is the $1$-Wasserstein distance between the two energy profiles $\mathcal P, \breve \mathcal P$, one might wonder whether ergotropy extraction protocols realise optimal transport plans in energy space. The answer is in the negative. 

In 1D the optimal transport plan $\bar \gamma$ that solves the optimisation problem in Eq. (\ref{eq:$1$-Wasserstein0}) is given by the Monge map $T=c_1^{-1}\circ c_2$, meaning that the two random variables $x,y$  are delta-correlated \cite{Vallender74TPA18}:
\begin{eqnarray}
\bar \gamma(x,y) = Q_2(y) \delta [x-c_1^{-1}(c_2(y))]
\end{eqnarray}
In our specific case it means that the optimal transport plan $\bar{\mathcal J}(E_1,E_2)$ reads
\begin{eqnarray}
\bar{\mathcal J}(E_1,E_2) = \breve{\mathcal P}(E_2) \delta [E_1-\mathcal C^{-1}(\breve{\mathcal C}(E_2))]
\end{eqnarray}
However, in the present case the two random quantities $E_1,E_2$ are generally not delta correlated. 
In the classical case, the joint probability density $\mathcal J(E_1,E_2)$ can be inferred from the conditional probability density $G[\Theta|\Phi]$, Eq. (\ref{eq:G}), that the system was on the energy hypersurface that encloses the volume $\Phi$ at the beginning, given that it is found on the energy hypersurface that encloses the volume $\Theta$ after ergotropy extraction. Thus the joint probability density of $\Theta,\Phi$ reads:
\begin{eqnarray}
J(\Theta,\Phi) =   \breve P(\Theta) G[\Theta|\Phi]
\end{eqnarray}
Note that its marginals are, correctly, $P(\Theta)$ and $\breve P (\Phi)$, see Eqs. (\ref{eq:P0-P1},\ref{eq:doubly-stoc}). Making the usual change of variables $\Phi =\Omega(E_1), \Theta = \Omega(E_2)$, we get:
\begin{eqnarray}
\mathcal J(E_2,E_1) = \breve{\mathcal P}(\Omega(E_2))   G[\Omega(E_1)|\Omega(E_2)]\omega(E_1)
\end{eqnarray}
where we note that $G$ is typically not a $\delta$ function (see \cite{Campisi08PRE78b} for an example). Thus  the optimal transport plan $\bar{\mathcal J} $ does not, in general, coincide with the actual ergotropy extraction ``plan'' ${\mathcal J} $. This is so because the ergotropic transport plan $\mathcal J$ is subject to the constraint that is induced by a symplectic diffeomorphism in phase space, whereas the optimisation in Eq. (\ref{eq:$1$-Wasserstein0}) is merely carried over all transport plans that are compatible with the marginals $\mathcal P$, $\breve{\mathcal P}$.

A similar argument applies as well in the quantum case.

\section*{Acknowledgements}
The sparkle that ignited the idea of this work was lit  at the ``Quantum Hike conference 2026'', Val D'Ayas, Italy. The author gratefully thanks the organisers (Sabrina Maniscalco, Lorenzo Maccone, Paola Verrucchi, Paolo Zanardi, Karol Z\.yczkowski) and the sponsor (Algorithmiq) for the opportunity to participate.
The author thanks Karen V. Hovhannisyan for his valuable remarks on the manuscript.
 
\section*{Use of artificial intelligence}
This work was solely produced by the author, without any aid of artificial intelligence.

\appendix
\section{Proof Eq. (\ref{eq:C<breveC}) in the quantum case}
In the quantum case the function $\mathcal C(E)$ is a histogram made of bars of variable widths and heights. The $i$-th bar has width $\Delta_i \equiv (E_{i+1}-E_{i})$ and height $s_i = \sum_{j=0}^i p_j$. The bars that compose the histogram $\breve \mathcal C(E)$ have the same widths $\Delta_i \equiv (E_{i+1}-E_{i})$ and their heigths are $\breve s_i = \sum_{j=1}^i r_j$. Because the overlaps $ |\langle{E_i}|r_j\rangle|^2$ form a bistochastic matrix, Eq. (\ref{eq:bistoch-quantum}) implies that the distribution $r_i$ majorises the distribution $p_i$ \cite{Sagawa20Book}, namely 
\begin{eqnarray}
\sum_{j=0}^i p_j^\downarrow \leq \sum_{j=0}^i r_j^\downarrow
\end{eqnarray}
where $p_j^\downarrow$ denotes the decreasing rearrangement of the distribution $p_j$. Note that by definition of decreasing rearrangement, it is  $ p_j \leq p_j^\downarrow$. Furthermore, from Eq. (\ref{eq:populations_ordered}) we have $ r_j^\downarrow=  r_j$. Therefore
\begin{eqnarray}
s_i =\sum_{j=0}^i p_j\leq  \sum_{j=0}^i   p_j^\downarrow \leq\sum_{j=0}^i r_j^\downarrow
= \sum_{j=0}^i r_j = \breve s_i
\end{eqnarray}
Thus each bar of the histogram of $\mathcal C$ is not higher than the corresponding bar of $\breve  \mathcal C$, meaning that $\mathcal C(E) \leq \breve  \mathcal C(E)$.

\section{Proof of Eq. (\ref{eq:D-qHO})}

Let
\begin{eqnarray}
f (x)\doteq \sum_{n=0}^{\infty} 
\left[
1- e^{-x} \sum_{k=0}^n  \frac{x^k}{k!}
\right] 
\label{eq:fx}
\end{eqnarray}
We want to prove that $f(x)=x$ for $x\geq 0$. In order to do so we first show that $f'(x)=1$. Noting that 
\begin{eqnarray}
e^{-x} \sum_{k=0}^\infty  \frac{x^k}{k!} =1
\end{eqnarray}
we can rewrite $f$ as:
\begin{eqnarray}
f(x)= e^{-x} \sum_{n=0}^\infty  \sum_{k=n+1}^\infty  \frac{x^k}{k!}
\end{eqnarray}
Let us take its derivative:
\begin{eqnarray}
f'(x)&= -e^{-x} \sum_{n=0}^\infty  \sum_{k=n+1}^\infty  \frac{x^k}{k!} +
e^{-x} \sum_{n=0}^\infty  \sum_{k=n+1}^\infty  \frac{x^{k-1}}{(k-1)!} \nonumber \\
&= -f(x) +e^{-x} \sum_{n=0}^\infty  \sum_{k=n}^\infty  \frac{x^k}{k!} \nonumber \\
&= -f(x) +e^{-x} \sum_{n=0}^\infty  \left [ \frac{x^n}{n!}+\sum_{k=n+1}^\infty  \frac{x^k}{k!}\right]  \nonumber\\
&= -f(x) +1 +f(x) \nonumber \\
&= 1
\end{eqnarray}
Furthermore, we have:
\begin{eqnarray}
\lim_{x\to 0} f(x) = 0
\end{eqnarray}
To prove it, note that $x^k/k!\to \delta_{0,k}$ for $x\to 0$, thus $e^{-x} \sum_{k=0}^n {x^k}/{k!} \to 1$ in Eq. (\ref{eq:fx}).

Integrating the equation $f'(s)=1$ between $0$ and $x$, with the condition $\lim_{x\to 0} f(x) =0$, we obtain $f(x)=x$.

\section*{References}

\bibliographystyle{iopart-num}

\begin{thebibliography}{10}
\expandafter\ifx\csname url\endcsname\relax
  \def\url#1{{\tt #1}}\fi
\expandafter\ifx\csname urlprefix\endcsname\relax\def\urlprefix{URL }\fi
\providecommand{\eprint}[2][]{\url{#2}}

\bibitem{Allahverdyan04EPL67}
{A E Allahverdyan}, {R Balian} and {Th M Nieuwenhuizen} 2004 {\em Europhys.
  Lett.\/} {\bf 67} 565--571

\bibitem{Campisi26EPL154}
Campisi M 2026 {\em Europhysics Letters\/} {\bf 154} 50002

\bibitem{Hatsopoulos76FP6b}
Hatsopoulos G~N and Gyftopoulos E~P 1976 {\em Found. Phys.\/} {\bf 6} 127--141

\bibitem{Gardner63PF6}
Gardner C~S 1963 {\em The Physics of Fluids\/} {\bf 6} 839--840
  \urlprefix\url{https://doi.org/10.1063/1.1706823}

\bibitem{Helander17JPP83}
Helander P 2017 {\em Journal of Plasma Physics\/} {\bf 83} 715830401

\bibitem{Lenard78JSP19}
Lenard A 1978 {\em J. Stat. Phys.\/} {\bf 19} 575--586

\bibitem{Pusz78CMP58}
Pusz W and Woronowicz S~L 1978 {\em Comm. Math. Phys.\/} {\bf 58} 273--290
  \urlprefix\url{https://doi.org/10.1007/BF01614224}

\bibitem{Gorecki80LMP4}
G{\'o}recki J and Pusz W 1980 {\em Letters in Mathematical Physics\/} {\bf 4}
  433--443 \urlprefix\url{https://doi.org/10.1007/BF00943428}

\bibitem{Daniels81JMP22}
Dani{\"e}ls H~A~M 1981 {\em J. Math. Phys.\/} {\bf 22} 843--846 ISSN 0022-2488
  \urlprefix\url{https://doi.org/10.1063/1.524949}

\bibitem{Mackenbach22PRL128}
Mackenbach R~J~J, Proll J~H~E and Helander P 2022 {\em Phys. Rev. Lett.\/} {\bf
  128} 175001

\bibitem{Lemou12IM187}
Lemou M, M{\'e}hats F and Rapha{\"e}l P 2012 {\em Inventiones mathematicae\/}
  {\bf 187} 145--194 \urlprefix\url{https://doi.org/10.1007/s00222-011-0332-9}

\bibitem{Campisi26arXiv:2603.28388}
{Campisi} M 2026 {\em arXiv:2603.28388\/}

\bibitem{Campisi25Book}
Campisi M 2025 {\em Lectures on the Mechanical Foundations of Thermodynamics\/}
  2nd ed SpringerBriefs in Physics (Cham: Springer)

\bibitem{Vallender74TPA18}
Vallender S~S 1974 {\em Theory of Probability and Its Applications\/} {\bf 18}
  784--786 \urlprefix\url{https://doi.org/10.1137/1118101}

\bibitem{Khinchin49Book}
Khinchin A 1949 {\em Mathematical Foundations of Statistical Mechanics\/} (New
  York: Dover)

\bibitem{Gradshteyn07Book}
IS~Gradshetyn I~R 1980 {\em Table of Integrals, Series, and Products\/}
  (Elsevier Academic Press)

\bibitem{Campisi08PRE78b}
Campisi M 2008 {\em Phys. Rev. E\/} {\bf 78} 051123

\bibitem{Sagawa20Book}
Sagawa T 2020 {\em Entropy, Divergence, and Majorization in Classical and
  Quantum Thermodynamics\/} SpringerBriefs in Mathematical Physics (Springer
  Singapore)

\end{thebibliography}

\providecommand{\newblock}{}

\end{document}